\documentclass[journal, a4paper]{IEEEtran}

\usepackage{slashed}
\usepackage{feynmp}
\usepackage{tikz-feynman}
\tikzfeynmanset{compat=1.1.0}
\usepackage{graphicx}
\usepackage{color}
\usepackage{amsmath,amssymb}   
\usepackage{subfigure} 
\usepackage{siunitx}
\usepackage[utf8]{inputenc}
\graphicspath{ {./images/} }
\usepackage{url}     
\begin{document}

\newcommand{\spr}[1]{{\color{red}\bf[SP:  {#1}]}}
\newcommand{\che}[1]{{\color{blue}\bf[CE:  {#1}]}}

	\title{\LARGE\bf Constraints on the maximal number of dark degrees of freedom\\ from black hole evaporation, cosmic rays, colliders, and supernovae}
	\author{Chris Ewasiuk and Stefano Profumo \\ 
 \textit{Department of Physics and Santa Cruz Institute for Particle Physics, \\
University of California, Santa Cruz, CA 95064, U.S.A.}}
	\maketitle

\begin{abstract}
	\textbf{A dark sector with a very large number of massive degrees of freedom is generically constrained by radiative corrections to Newton's constant. However, there are caveats to this statement, especially if the degrees of freedom are light or mass-less. Here, we examine in detail and update a number of constraints on the possible number of dark degrees of freedom, including from black hole evaporation, from perturbations to systems including an evaporating black hole, from direct gravitational production at colliders, from high-energy cosmic rays, and from supernovae energy losses.}
\end{abstract}

\section{Introduction}
Dark sectors -- extensions to the field content of the Standard Model  (SM) of particle physics containing degrees of freedom generally very weakly coupled, or exclusively gravitationally coupled, to SM degrees of freedom --  can generically be very large. Twin Higgs models \cite{Chacko:2005pe}, supersymmetric extensions to the SM \cite{Martin:1997ns}, and mirror models \cite{Foot:1991bp} all have comparable numbers of degrees of freedom as the SM itself. Models with large extra dimensions can have extremely large numbers of degrees of freedom, from the perspective of the effective four-dimensional theory: indicating with $M_*$ the $d$-dimensional Planck mass, the effective 4 dimensional Planck scale is given by
\begin{equation}
M_P^2=M_*^{d-2}V_{d-4},
\end{equation}
where $V_{d-4}$ is the volume of the extra dimensions. If the latter is large, as envisioned for instance in the realizations of Ref.~\cite{Arkani-Hamed:1998jmv, Hooper:2007qk}, the four-dimensional theory  contains a potentially extremely large number of degrees of freedom, in fact on the order of $M_*^{d-4}V_{d-4}$. In addition to similar extra-dimensional setups, other examples of scenarios containing large dark sectors are, for instance, the  ``clockwork''  \cite{Giudice:2016yja} and ``stasis'' \cite{Dienes:2021woi} scenarios.

Even absent any interaction ``portal'' with the SM, a large number of degrees of freedom entails, in turn, radiative corrections to the gravitational coupling $G(\mu=0)=G_N$ that affect, through renormalization group running, that constant's value as a function of the energy scale $\mu$. In particular, for light, but massive, degrees of freedom with mass $m$ lighter than $\mu$, Ref.~\cite{ref17} computes that the correction to $G_N$ would read:
\begin{equation}\label{eq:mdep}
\delta G^{-1}(\mu)=G^{-1}(\mu)-G^{-1}(0)\sim N m^2\ln\left(\mu^2/m^2\right),\  (m<\mu).
\end{equation}
The most conservative constraints, which entail, that is, the least model-specific assumptions from precision gravity tests \cite{ref47, ref48} are that 
\begin{equation}
\frac{\delta G}{G}\lesssim 10^{-9}
\end{equation}
at distances on the order of $d\sim 10^3$ km, corresponding to energy scales (in natural units, assumed everywhere here) $\mu\sim 1/d\sim 10^{-13}$ eV. In turn, since
\begin{equation}
\frac{\delta G}{G}\sim N (m/M_P)^2,
\end{equation}
precision gravity constraints imply that
\begin{equation}
N\lesssim 10^{-9}\left(\frac{10^{-13}\ {\rm eV}}{M_P}\right)^{-2}\sim 10^{73}.
\end{equation}
A similar conclusion follows from a different approach, but, again, based on renormalization effects on Newton's constant, presented in \cite{Calmet}, that finds that for any light degree of freedom $m\lesssim\mu$, 
\begin{equation}
\delta G^{-1}(\mu)=G^{-1}(\mu)-G^{-1}(0)\sim N \mu^2/(12\pi).
\end{equation}
Precision gravity tests here as well imply the weak limit $N\lesssim 10^{73}$. 

A qualitatively different situation occurs considering the implications of strong gravity at scales close to those at which the SM is tested with great precision, $\mu\lesssim 1$ TeV. Such considerations, presented e.g. in \cite{ref17, ref19}, assume that the running of $G$ can be calculated perturbatively, an assumption that certainly breaks down in the regime of strong gravity, and additionally assume that no other effects perturb $G(\mu)$. For instance, while Ref.~\cite{Calmet,Pal:2019tqq} point out that scalar and Weyl fermions add coherently to the renormalization of $G(\mu)$, gauge bosons contribute with the opposite sign. Since the cosmological constant would also receive radiative contributions from such degrees of freedom, large cancellations would not be entirely out of question. Finally, if the running depends on the new degrees of freedom's mass $m$ as in Eq.~(\ref{eq:mdep}), requiring strong gravity to occur at scales greater than $\mu_*\gtrsim 1$ TeV implies
\begin{equation}
N\lesssim \frac{(M_P/m^2)^2}{\ln(\mu_*^2/m^2)}\simeq 10^{78}\left(\frac{10^{-12}\ {\rm eV}}{m}\right)^2.
\end{equation}
and would thus be rather weak in the case of light dark sectors. 

By far the most stringent, quasi-model-independent constraints on  large numbers of gravitationally-coupled dark degrees of freedom is expected to stem from the existence, in the late universe, of stellar-mass black holes. Black holes are known to evaporate via quantum fluctuations around the event horizon in a process known as Hawking radiation. Gibbons and Hawking showed that black holes emit quasi-thermal radiation at a temperature

\begin{equation}
T_H = \frac{1}{8 \pi M },
\end{equation}
where $M$ is the black hole mass \cite{Hawking,BH}. This temperature acts effectively as an upper threshold to the mass of the particles that could be produced at the event horizon: any particle with rest mass energy much larger than $T_H$ will be emitted at a Boltzmann exponentially-suppressed rate. Since the temperature of the black hole is, in turn, inversely proportional to its mass, the majority of SM degrees of freedom are not be emitted until the last moments of a massive black hole's life cycle.

While black hole emission is close to thermal, significant effects on the particle emission rates are related to the black hole's spin and charge(s), as well as to the charge, spin, and mass of the radiated particles \cite{Page:1976ki}. 
Such effects are encapsulated in the so-called grey-body factors that we discuss below. In particular, in what follows we study in detail how the mass and spin of the light dark sector particles, as well as the mass and spin of the radiating black hole, affect the hole's lifetime, and explore in detail both the black hole merger systems providing the strongest constraints on the number of dark degrees of freedom, and the dependence of such constraints on the model-specific variables of the dark sector under consideration.

While it is reasonable to assume that stellar-mass black holes observed to have a certain mass today were produced with a comparable mass, the {\it initial} black hole mass is, in principle, unknown. Evaporation constraints based on the currently observed (or inferred) black hole mass are thus subject to the caveat that the same black holes be born with a much larger mass, and have evaporated away to the current mass. In this case, constraints can only be derived from {\it direct} effects of black hole evaporation on systems containing a black hole. Such systems could be significantly perturbed by the evaporation, and associated mass-loss process, in ways that are in principle detectable or observable. We discuss this possibility as well in what follows.

Other phenomenological constraints on the number of dark degrees of freedom the present study will focus on stem from: 

(1) gravitational collider production of the dark states and searches for the resulting $2\to3$ process involving the pair production of invisible dark states and initial state strongly-interacting radiation (manifesting as jets); 

(2) production of dark sector particles at very high-energy cosmic-ray collisions, and 

(3) cooling of supernovae by gravitationally produced dark sector particles.  

All such constraints were, schematically, considered in \cite{hooman}, which, however, only carried out an order of magnitude estimation. We here carry out a detailed computation of the relevant cross sections, accurate numerical integrations, and obtain results that in part confirm and in part are weaker, or stronger than those quoted in Ref.~\cite{hooman}. In all cases, however, we find that the constraints from black hole evaporation -- albeit not with the caveat discussed above -- are stronger.

Finally, gravitational particle production in the early universe can additionally constrain large dark sectors, by over-closing the universe; we also discuss below model-specific instances of such constraints, and relevant caveats.

The remainder of this paper is structured as follows. In the following section we discuss constraints on the number of dark sector degrees of freedom from the existence of stellar-mass black holes in the late universe; as alluded to above we also entertain possible caveats to such constraints, and derive more conservative constraints based on direct evidence of evaporation in the perturbation of systems containing a black hole (sec.~\ref{sec:caveats}). We then re-assess and update, in the following sec.~\ref{sec:colliders} the constraints from gravitational production of dark sector particles at very high-energy cosmic ray events, at colliders, and at supernova explosions; we discuss gravitational particle production in sec.~\ref{sec:cosmo}, and, finally, present a discussion and our conclusions in the closing sec.~\ref{sec:conclusions}.

\section{Constraints from late-universe stellar-mass black holes}
A consequence of introducing extra dark degrees of freedom is the increased evaporation rate of black holes (BHs) through Hawking radiation \cite{Hawking}. Observational data on BH mergers from gravitational wave observatories offer insights into candidates that provide constraints on the ``size'' of dark sectors.  
\subsection{Constraints from gravitational wave events}
As detailed above, BHs radiate, approximately,  as black bodies; Since the holes' temperature is inversely proportional to the area enclosed by the event horizon, whose radius is in turn proportional to the hole's mass, the Stefan-Boltzmann law then dictates that mass loss be inversely proportional to the square of the mass. Thus the {\it lifetime} of a BH of mass $M$ scales, approximately and in natural units, as $\tau\sim (1/N) (M^3/M_{\rm Pl}^4)$, with $N$ the number of degrees of freedom.  The constraints on $N$ here can thus be estimated as
\begin{equation}\label{eq:roughconst}
N<\frac{({M}^3/\tau)}{M^4_{\rm Pl}}.
\end{equation}
As a result, the most constraining BHs are the {\it lightest} and {\it oldest} firmly detected BHs, thus minimizing the ratio $M^3/\tau$. Supermassive BHs, and BHs associated with X-ray binaries are significantly more massive than the lightest BHs detected by LIGO via the gravitational waves emitted upon their merger. We will discuss below, however, caveats to the constraint outlined in Eq.~(\ref{eq:roughconst}) where the former objects will become highly relevant.

\begin{figure*}[t!]
\begin{center}
\includegraphics[width = \textwidth, height = 0.45\textwidth ]{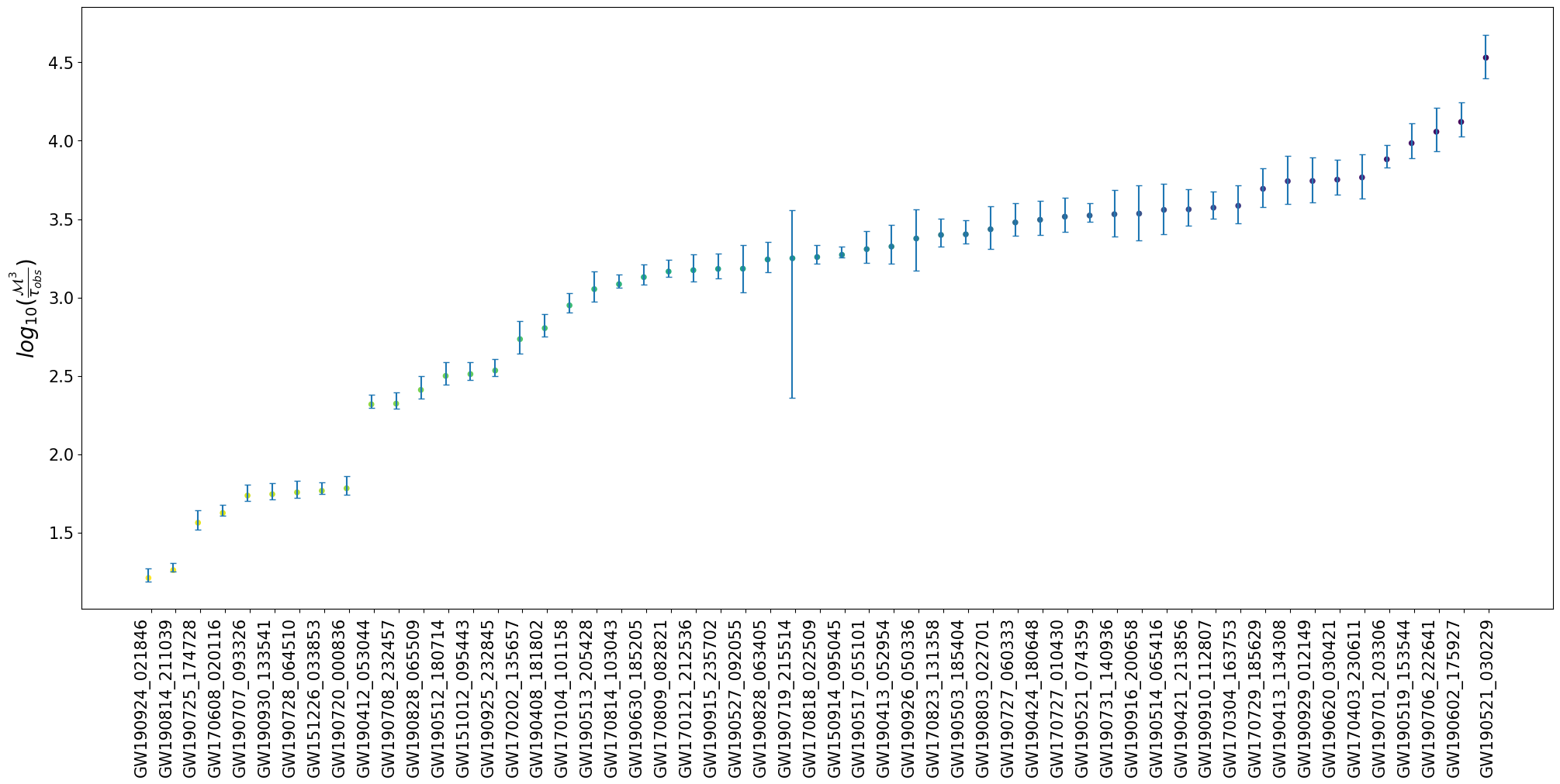}
\caption{Histogram illustrating LIGO/VIRGO candidates for constraints on the number of large dark sector degrees of freedom. For each event, we list inferred values of $M^3/\tau_{\rm obs}$ and the corresponding, propagated errorbars. Data  from Ref.~\cite{Ligo}. }
\label{fig:Hist}
\end{center}
\end{figure*}

LIGO indirectly measures, or allows to infer, the mass of the merging BHs and, in some cases, 
the distance from the detector of the gravitational waves' source via their waveform amplitude \cite{BHGW150914}. Redshift information is then extracted by either the measured distribution of masses, or 
via the distribution of redshifts mapped by a galaxy catalogue, such as GLADE+ \cite{BHGW150914}. 
Since LIGO's stellar-mass BHs are believed to have formed no later than redshift $z_{\rm form} \sim 10$ \cite{Formr}, using both the redshift and the estimated formation time, we can obtain a proxy for an upper limit on $\tau$, i.e. how long the BH has been, albeit indirectly through the merger event, observed to have lived. 

We list all candidates for constraints, from the merger catalogue of Ref.~\cite{Ligo} in Figure \ref{fig:Hist}, as a histogram of the ratio of $ \mathcal{M}^3 / \tau_{\rm obs}$, where $\mathcal{M}$ refers to the chirp mass, defined as
\begin{equation}
    \mathcal{M} = \frac{(m_1m_2)^{3/5}}{(m_1+m_2)^{1/5}}.
\end{equation}
The chirp mass is a combination of the inspiraling BH masses  and is the relevant parameter for determining the inspiral evolution of a merger event.

The error associated with the ratio $\mathcal{M}^3/\tau_{\rm obs}$ is propagated through the uncertainties in the chirp mass and observed lifetime, where the upper and lower uncertainty bounds are averaged to obtain the error associated with fig. \ref{fig:Hist}.  Conveniently, and interestingly, the best-constraining candidate event also has the second lowest propagated uncertainty associated with its mass and redshift measurements. Candidate merger events with individual masses equal to or lower than the Tolman–Oppenheimer–Volkof limit \cite{Kalogera:1996ci} were neglected, as these merger events likely contain neutron stars rather than BHs. 
As mentioned above, this quantity is a proxy for the constraints we are interested in. 

From the list shown in fig.~\ref{fig:Hist}, we infer that the two most-constraining events correspond to gravitational wave GW190924\_021846 and GW190814\_211039. However, one of the masses involved in event GW190814\_211039 lies at the edge of the TOV limit \cite{Kalogera:1996ci}, so we  exclude this event from our analysis. Therefore, in what follows we will use values corresponding to GW190924\_021846,  with an observed lifetime of $\tau_{\rm obs} \simeq 1.157^{+0.048}_{-0.034} \times 10^{10}$ years, chirp mass $\mathcal{M} = 5.7^{+0.2}_{-0.1} M_{\odot}$, and effective spin parameter  $a_{\rm obs}^* = 0.05^{+0.16}_{-0.1}$, defined as 
\begin{equation}
a_{\rm obs} ^*= \frac{m_1 a_{1}^*\cos{\theta_1} + m_2 a_{2}^*\cos{\theta_2}}{m_1+m_2},
\end{equation}
where the dimensionless angular momentum per mass parameter is defined as $a^*\equiv J/M^2$.
The effective spin parameter takes into account the individual masses and spin parameters $m_i$ and $a^*_i$ of the BH merger, as well as the angle $\theta_i = \cos^{-1}{(L \cdot S_i)}$ between the individual BH spin $S_i$ and the orbital angular momentum vector $L$ \cite{Fernandez_2019}. 
The BH mass used in the \textit{BlackHawk} computations below is taken to be the smaller of the two masses for our ideal merger event. This corresponds to $M=4.8 M_{\odot}$ for GW19092\_021846. The uncertainties associated with the relevant merger event parameters extracted from the LIGO catalogue are taken to be approximately symmetric and Gaussian in origin and are illustrated in fig. \ref{fig:Hist}. Additionally, for definiteness and simplicity, the BH is taken to be rotating with a spin parameter equivalent to the effective spin of the orbital system. In the case of GW19092\_021846 this corresponds to $a^*=0.05$.

\begin{figure}[b!]
\begin{center}\includegraphics[width = .5\textwidth, height = .45\textwidth]{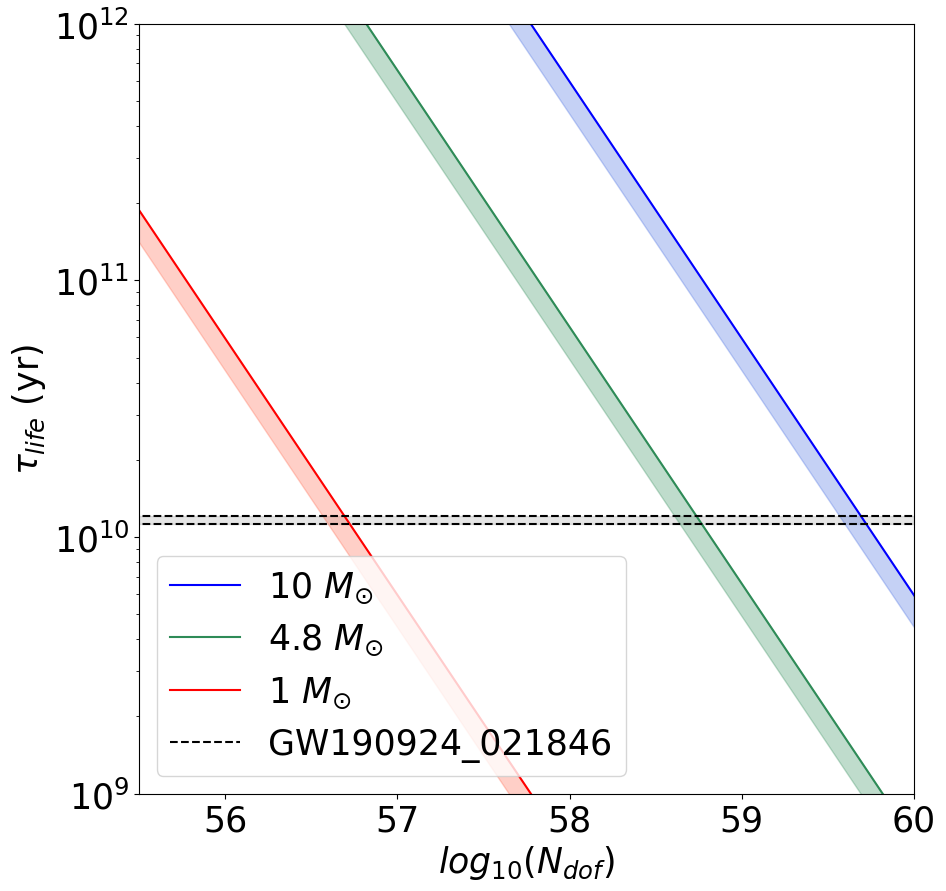}
\caption{Effect of a large number of dark degrees of freedom on the lifetime of a BH as a function of the number of dark degrees of freedom, for three masses, 1 (red), 4.8 (corresponding to the merger event  GW190924\_021846) and 10 $M_\odot$. The solid lines correspond to $a^* = 0$ and the corresponding shaded area extends down until it approaches a maximally spinning BH, $a^* \rightarrow 1$. The dashed lines and grey area correspond to the observed lifetime of BH merger event  GW190924\_021846, $\tau_{obs} = 1.157^{+0.048}_{-0.034} \times 10^{10}$ years.} 
\label{fig:LTvsDOF}
\end{center}
\end{figure}

\subsection{Constraints on dark sectors degrees of freedom from LIGO merger detections}
Constraints on the number of dark sector degrees of freedom depend on both the mass spectrum and spin of such degrees of freedom, and on the spin (and charge, in principle, although astrophysical BHs would discharge extremely rapidly \cite{discharge}) of the evaporating BH. 

BHs evaporation rates and the greybody factors -- encapsulating the deviation from a perfect black body -- depend both on the mass and on the dimensionless angular momentum per mass parameter $a^*$. For a  BH of mass $M$ and  spin parameter $a^*$, the so-called Kerr horizon \cite{Kerr} is given by
\begin{equation}
    r_{H } = r_{S} \frac{1 + \sqrt{1 -a^{*2}}}{2}
\end{equation}
For $a^* \rightarrow 0$ one of course recovers the static, Schwarzschild solution, while $a^* \rightarrow 1$ corresponds to the extremal solution where $r_H\to0$.  
In turn, the corresponding Kerr BH temperature is given by
\begin{equation}
    T_K = \frac{1}{2 \pi} \frac{r_+ - M}{r_+^2 + a^{*2}},
\end{equation}
and the emission rates for a given degree of freedom $i$ per unit time and energy is given by the expression \cite{BH}
\begin{equation}\label{eq:grebody}
    \frac{d^2N_{i,lm}}{dtdE} = \frac{1}{2 \pi} \frac{\Gamma_{s_i,lm}(E,M,a^*)}{e^{E/T_K} - (-1)^{2s_i}},
\end{equation}
where $\Gamma_{s_i,lm}$ is the greybody factor, which encodes the  probability for a particle of spin $s_i$, with  to be created on the horizon and escape  to infinity; in the expression above, $lm$ indicate the particle angular momentum $l\ge s_i$, and its projection on the BH axis of rotation $m =-l, ... , +l$. 

We first consider a dark sector consisting of a collection of {\it mass-less} and {\it spin-less} (i.e. spin 0) scalars. Using the state-of-the-art \textit{Blackhawk} code \cite{BH}, we  calculate the BH lifetime as a function of the number of dark sector degrees of freedom for BH masses corresponding, for comparison, to $M=1,\ 4.8$ and 10 $M_\odot$. We show our results in fig.~\ref{fig:LTvsDOF}. In the figure, the shading shows the  effect of BH spin on the lifetime, with the solid line corresponding to $a^*=0$ and the lower boundary of the shaded region to $a^*=1$: increasing the BH spin  generally speeds up the rate of evaporation of the BH, as explained above.

As the figure shows,  the effect of the BH's angular momentum is rather small. Since the dark degrees of freedom dominate all other modes of particle production, one finds an approximately linear relationship between lifetime and PBH mass, stemming from the equation \cite{BH}
\begin{equation}
    \frac{dM_{BH}}{dt} = - \sum_i \frac{f_i(M_{BH}, a^*)}{M_{BH}^2} \approx \frac{f_{\rm DS}(M_{BH}, a^*)}{M_{BH}^2}.
    \label{MASSLOSS}
\end{equation}

The above mass loss rate is dominated by $f_{\rm DS}(M_{BH}, a^*)$, which in turn is directly proportional to the degrees of freedom of the dark sector $N_{\rm dof}$.  The estimated lifetime of the PBH is derived under the assumption that the BH formed at a redshift of $z \approx 10$  (0.478 Gyr after the big bang) , and exists until its merger at the redshift listed within the LIGO catalogue \cite{Ligo}. The degrees of freedom for a given BH mass were then chosen to match the inferred lifetime of the PBH, as shown in fig. \ref{fig:LTvsDOF}.

\begin{figure}[t!]
\begin{center}
\includegraphics[width = .5\textwidth, height = .45\textwidth]{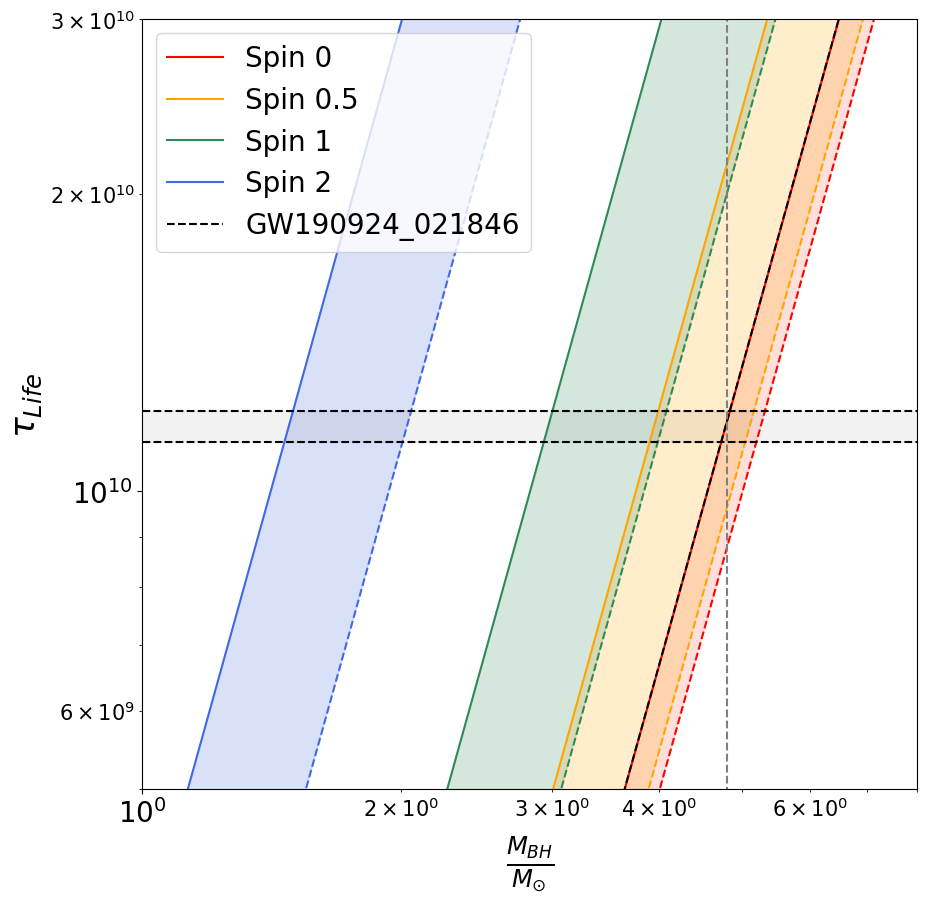}
\caption{ Effect of the dark sector particles' spin on the lifetime, in years, of a BH of different intrinsic spin. Solid colored lines represent evaporating BHs with spin $a^* = 0$, while dashed lines to a maximally spinning BH ($a^* \rightarrow 1$). The BH mass is given in terms of solar masses, and the vertical line corresponds to the mass of the lightest BH in the GW19092\_021846 event. We assume $5 \times 10^{58}$  dark degrees of freedom and  massless dark sector particles, $M_{\rm DS} = 0$ GeV, with spin 0 (red), 1/2 (orange), 1 (green), and 2 (blue). } 
\label{fig:LTvsMBH}
\end{center}
\end{figure}

We now turn to the effects of the dark sector's particles spin. As explained above, the effect of the particle's spin is encoded in the greybody factors of Eq.~(\ref{eq:grebody}). Figure \ref{fig:LTvsMBH} shows constraints on $N_{\rm dof}$ determined by comparing the observed BH lifetime to the lifetime corresponding to a given number of dark degrees of freedom, here for illustration purposes set to $N_{\rm dof}\simeq 5 \times 10^{58}$. In the figure, the solid lines corresponds to an evaporating Schwarzschild BH ($a^*=0$), while the dashed lines to an extremal Kerr BH ($a^*=1$), and the different shaded areas show results for spin 0 (red), 1/2 (orange), 1 (green), and 2 (blue). We also show with a grey, horizontal band the lifetime associated with the GW190924\_021846 event $1.157^{+0.048}_{-0.034} \times 10^{10}$. 

The figure demonstrates that dark particles with lower spin will result in shorter  lifetimes (although this depends on the BH's assumed spin). In other words, dark sector particles with lower spin will require fewer degrees of freedom than higher spin particles to produce the same BH lifetime. We note that for BHs with larger spin the rate of emission of particles with larger spins is larger \cite{Arbey:2021mbl}.

\begin{figure}[t!]
\begin{center}
    
\includegraphics[width = .5\textwidth, height = .45\textwidth]{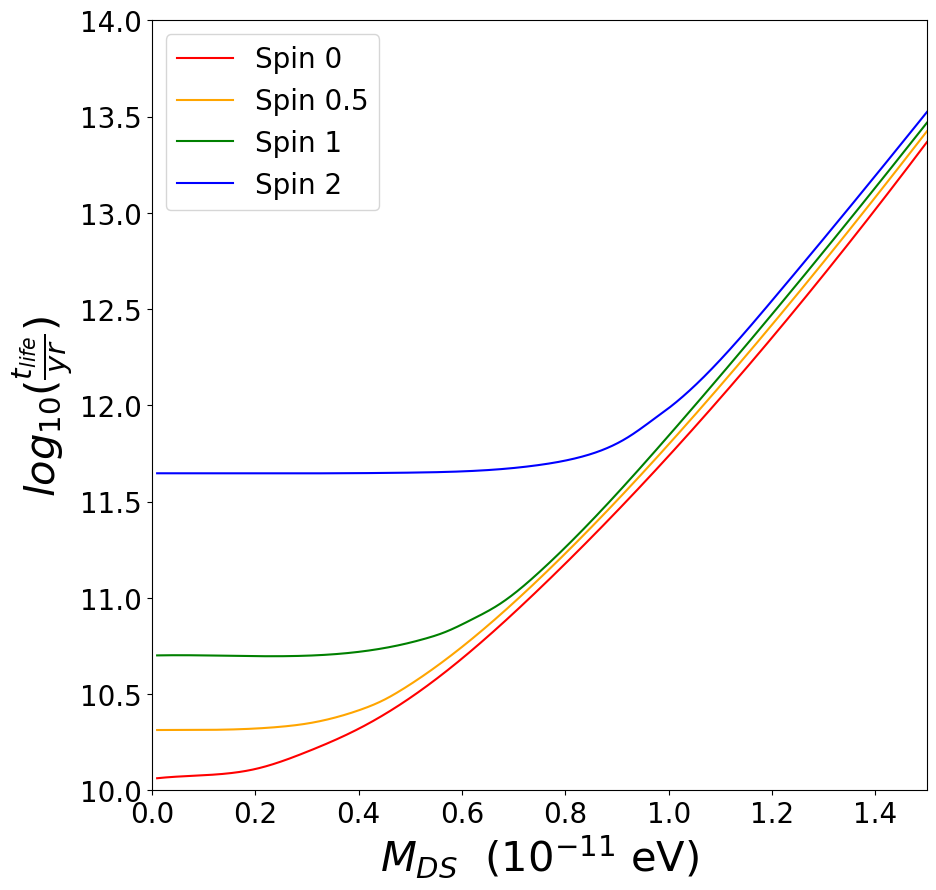}
\caption{Effect of dark sector mass and particle spin on the evaporating BH lifetime, for dark sectors of different spin and mass. The BH is initially assumed to have mass $M = 4.8 M_{\odot}$, $a^* = 0.05$, and we assume a number of degrees of freedom $N_{\rm dof} = 5 \times 10^{58}$.}
\label{fig:LTvsMDS}
\end{center}
\end{figure}

Next, we consider the case where we vary the mass-scale of the proposed dark sector, with all other parameters held constant. We assume that all dark sector degrees of freedom be degenerate in mass at a mass-scale $M_{S}$. The results are presented in fig.~\ref{fig:LTvsMDS}. Again we observe the pattern of lower spins radiating more efficiently, lowering the BH lifetime. The evaporation into particles with rest masses greater than those allowed by the initial temperature of the BH is exponentially suppressed. The only particles able to be copiously radiated are those with a mass comparable to the BH's temperature. The effect approaches a linear relationship for rest masses greater than $ 1 \times 10^{-20}$ GeV, for solar-mass BHs. The linear relationship continues until the lifetime of the BH matches that of the SM, in which case the threshold mass of the proposed particle will never be reached. This occurs initially around $4.5 \times 10^{-21} $ GeV where the spin 0.5 particle begins to radiate more efficiently than the spin 0 particle. This only occurs for a limited mass range, as spin 0 begins to radiate more efficiently for masses larger than $5.5 \times 10 ^{-21}$.

\begin{figure}[t!]
\begin{center}    
\includegraphics[width = .5\textwidth, height = .45\textwidth]{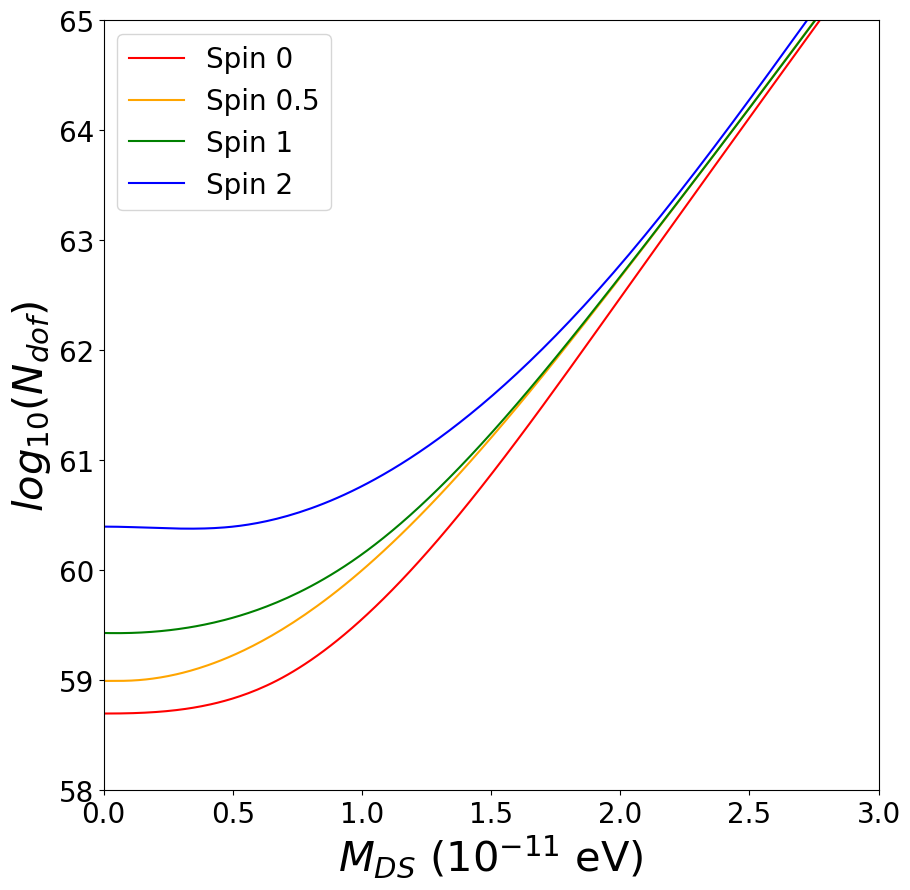}
\caption{Effect of dark sector mass and particle spin on the number of degrees of freedom corresponding to the observed lifetime of $\tau_{obs} = 1.157^{+0.048}_{-0.034} \times 10^{10}$ years of the BHs in GW190924\_021846.  The initial BH parameters are $M = 4.8 M_{\odot}$ and  $a^* = 0.05$.}
\label{fig:NvsMDS}
\end{center}
\end{figure}

In our final figure \ref{fig:NvsMDS}, we summarize the constraints from the BH lifetime corresponding to the mass and spin of GW190924\_021846 on the plane defined by the number of degrees of freedom $N_{\rm dof}$ and the dark sector mass scale $M_{\rm DS}$, for dark sectors with spin 0, 1/2, 1 and 2 (respectively red, orange, green, and blue lines). Points {\it above} the lines are excluded. Again, given the low effective spin of GW190924\_021846, the strongest constraints stem from $s=0$ dark sectors; as expected, constraints from evaporation weaken significantly for dark sectors above $M_{\rm DS}\sim M_{\rm Pl}^2/M_{\rm BH}\sim 10^{-11}$ eV.

In conclusion, using the {\it most constraining} known black hole, i.e. the longest-lived and the lightest one, we were able to constrain the size of a {\it massless} dark sector to between $N_{s=2}= 2.5 \times 10 ^{60}$  and $N_0= 5 \times 10^{58}$; for massive dark sectors, the detailed results are shown in fig. \ref{fig:NvsMDS}.

\subsection{Caveats to Black Hole Lifetime constraints}\label{sec:caveats}
A caveat to the constraints discussed above is that BHs may be {\it born} more massive, undergo evaporation, and end up with the mass they are observed to have at the moment of the binary merger. Albeit fine-tuned, this scenario is worth considering.

How can one then constrain large evaporation rates with available observations? We identify three possible routes:
\begin{enumerate}
    \item Evaporation would lead to an additional energy-loss process competing with gravitational wave emission upon binary mergers. If the mass-loss process due to evaporation $\dot E_{\rm EV}$ consumes a comparable amount of power as gravitational wave emission $\dot E_{\rm GW}$, i.e. $\dot E_{\rm EV}\sim\dot E_{\rm GW}$, the merger development would measurably change, and give a measurably different pattern of gravitational wave emission, or even not occur at all. Conservatively, we then require $\dot E_{\rm EV}<\dot E_{\rm GW}$;
    
    \item Assuming that X-ray binaries that include a BH emit, in their quiescent state, proportionally to the Eddington rate, which in turn is proportional to the BH's mass, the luminosity change $\Delta L/L$ over a time interval $\Delta t$ would then be correlated with the BH's mass change, i.e. $\Delta m/m\sim \Delta L/L$, and 
    \begin{equation}
       \frac{1}{M} \frac{{\rm d}M}{{\rm d}t}\equiv\frac{\dot M}{M}\sim\frac{\Delta M}{M\Delta t}\lesssim \frac{\Delta L}{L}.
    \end{equation}    
    \item More generally, one can utilize BH mass determination $M_{\rm BH}=m\pm\Delta m$, from a variety of observations taken over a time span $\Delta t$, and constrain 
    \begin{equation}
      \frac{\dot M}{M}\lesssim \frac{\Delta m}{m}\frac{1}{\Delta t}.
    \end{equation}   
\end{enumerate}

For the first constraint, we assume, for simplicity an equal mass BH binary merger with BH masses $M$ on a circular orbit of radius $R$. The energy loss to gravitational wave then reads \cite{carroll}:
\begin{equation}
    \frac{{\rm d}E_{\rm GW}}{{\rm d}t}=-\frac{2}{5}\frac{G_N^4}{c^5}\frac{M^5}{R^5};
\end{equation}
the gravitational energy loss due to the mass loss reads, instead,
\begin{equation}
    \frac{{\rm d}U}{{\rm d}t}=\frac{\rm d}{{\rm d}t}\left(-\frac{G_NM^2}{R}\right)=-\frac{2M^2G_N}{R}\frac{\dot M}{M}.
\end{equation}
Using the two expressions above, we find
\begin{equation}
    \frac{\dot M}{M}<\frac{G_N^3M^3}{5R^4}. 
\end{equation}
Using the relation between angular frequency $\Omega$, radius and mass
\begin{equation}
    G_N M=4R^3\Omega^2,
\end{equation}
we find
\begin{equation}
    \frac{\dot M}{M}<\frac{\Omega^{8/3}}{c^5}\left(G_N M\right)^{5/3}. 
\end{equation}
Using the equation above, for $M\sim 10\ M_\odot$ and $\Omega\sim 10^3$ Hz, we find that 
\begin{equation}
    \left(\frac{\dot M}{M}\right)_{\rm GW}\lesssim 10\ {\rm Hz}.
\end{equation}

As for the second constraint, utilizing the data shown in Tab. 2 of Ref.~\cite{2009ApJS..181..238C}, where, in a quiescent state BW Cir is observed to have an approximately constant luminosity, within $\sim10\%$ errorbars, over a time interval of 14 years, or $\Delta t
\sim 4\times 10^8$ sec, we can constrain
\begin{equation}
    \left(\frac{\dot M}{M}\right)_{\rm lum}\lesssim\frac{\Delta L}{L}\frac{1}{\Delta t}\sim 2\times 10^{-10}\ {\rm Hz},
\end{equation}
thus over 10 orders of magnitude more stringent than the constraint discussed above, but for a larger mass.

In the last approach, we consider a list of all known X-ray binary systems, from Ref.~ \cite{Filippenko:1999zv, Plotkin:2021rzl}, both Galactic and extragalactic, including from gravitational wave observations as listed above. Using as a criterion the smallness of $\Delta m/m$, the best candidates are Sag A*, whose mass is known to better than 0.3\%, and the Gaia BH1 and GRS 1009-45 /
Nova Velorum 1993/MM Velorum systems, whose masses are known to better than 3\%. The latter two, however, are significantly lighter BHs, even though the observation time is shorter than for Sag A*. We obtain, using the central value for the mass of the BH in the Nova Velorum 1993 system of $4.3\ M_\odot$ and one year of observation time \cite{Filippenko:1999zv},
\begin{equation}
    \left(\frac{\dot M}{M}\right)_{\rm XRB}\lesssim 10^{-9}\ {\rm Hz}.
\end{equation}
For Sag A* we find instead, assuming 15 years of exposure,
\begin{equation}
    \left(\frac{\dot M}{M}\right)_{\rm SagA*}\lesssim 6\times 10^{-12}\ {\rm Hz}.
\end{equation}

How do these constraints compare to constraints from complete evaporation? Let us recall that, from
\begin{equation}
    \frac{dM}{dt}=-\frac{\alpha}{M^2},
\end{equation}
the lifetime, assuming constant $\alpha$, is $\tau=M^3/(3\alpha)$. The constraints from evaporation (translating as described above into limits on the number of additional dark degrees of freedom) thus scale with the evaporation rate $\alpha$ as
\begin{equation}
    \alpha_{\rm ev}<\frac{M^3}{3\tau}.
\end{equation}
The constraints outlined above instead read
\begin{equation}
    \alpha_{\rm lim} \sim M^3 \left(\frac{\dot M}{M}\right)_{\rm lim},
\end{equation}
the latter quantity having been calculated for the three methods outlined above. Using, for reference, $M_{\rm ev}=5.7\ M_\odot$, $N_{\rm ev}=3\times 10^{59}$ and $\tau_{\rm ev}=1.34\times 10^{10}$ yr, the estimate of the constraint to the number of dark degrees of freedom compared to that from evaporation, is given by
\begin{eqnarray}\nonumber
    \frac{N}{N_{\rm ev}}&\lesssim& \frac{M^3\left(\frac{\dot M}{M}\right)_{\rm lim}}{\alpha_{\rm ev}}=\frac{M^3\left(\frac{\dot M}{M}\right)_{\rm lim}}{M_{\rm ev}^3/(3\tau_{\rm ev})}\\[0.2cm]
    &&=3\tau_{\rm ev}\left(\frac{M}{M_{\rm ev}}\right)^3\left(\frac{\dot M}{M}\right)_{\rm lim}.
\end{eqnarray}

As a result, we get that the GW constraints give
\begin{equation}
    \frac{N_{\rm GW}}{N_{\rm ev}}\sim 7\times 10^{19},\ \ {\rm or}\ \ N_{\rm GW}<2\times 10^{79};
\end{equation}
the constraints from BW Cir (assuming a mass of 7 $M_\odot$ \cite{2009ApJS..181..238C}) give
\begin{equation}
    \frac{N_{\rm BWCir}}{N_{\rm ev}}\sim 5\times 10^{8},\ \ {\rm or}\ \ N_{\rm BWCir}<10^{68};
\end{equation}
the constraints from XRB GRS 1009-45 \cite{Filippenko:1999zv, Plotkin:2021rzl} give
\begin{equation}
    \frac{N_{\rm XRB}}{N_{\rm ev}}\sim 5\times 10^8,\ \ {\rm or}\ \ N_{\rm XRB}<2\times 10^{68}.
\end{equation}
Finally the constraints from Sag A* (15 year time scale, mass of $4.3\times 10^6\ M_\odot$ \cite{GRAVITY:2021xju}) give
\begin{equation}
    \frac{N_{\rm SagA*}}{N_{\rm ev}}\sim 3\times 10^{24},\ \ {\rm or}\ \ N_{\rm SagA*}<10^{82}.
\end{equation}

One can also wonder whether the fact that X-ray binaries exhibit a large spin parameter $a^*$ could lead to stronger constraints. To this end, consider the time evolution of the spin parameter \cite{Page:1976ki}
\begin{equation}
    \frac{da^*}{dt}=a^*\frac{2f(M_{BH},a^*)-g(M_{BH},a^*)}{M^3},
\end{equation}
where $g$ is a function analog to $f$, and both $f,\ g\sim N_{\rm dof}$. Substituting the expression for the mass evolution, and assuming both $f$ and $g$ constant for simplicity, we find that 
\begin{equation}
    a(t)\simeq a^*(t_0)e^{-\frac{2f-g}{3f}}\left(\frac{M(t)}{M(t_0)}\right)^3,
\end{equation}
where $t_0$ is the BH formation time. We thus find that the timescale for a significant reduction in the intrinsic spin parameter is similar to that for the mass. We thus conclude that the order of magnitude estimate derived above for mass also holds even considering deviations from $a^*\simeq 1$.

\section{High-Energy Cosmic Ray, Collider, and Supernova constraints}\label{sec:colliders}

\subsection{Ultrahigh Energy Cosmic Rays}
\begin{figure}[!t]
    \centering
\includegraphics[width = 0.45\textwidth]{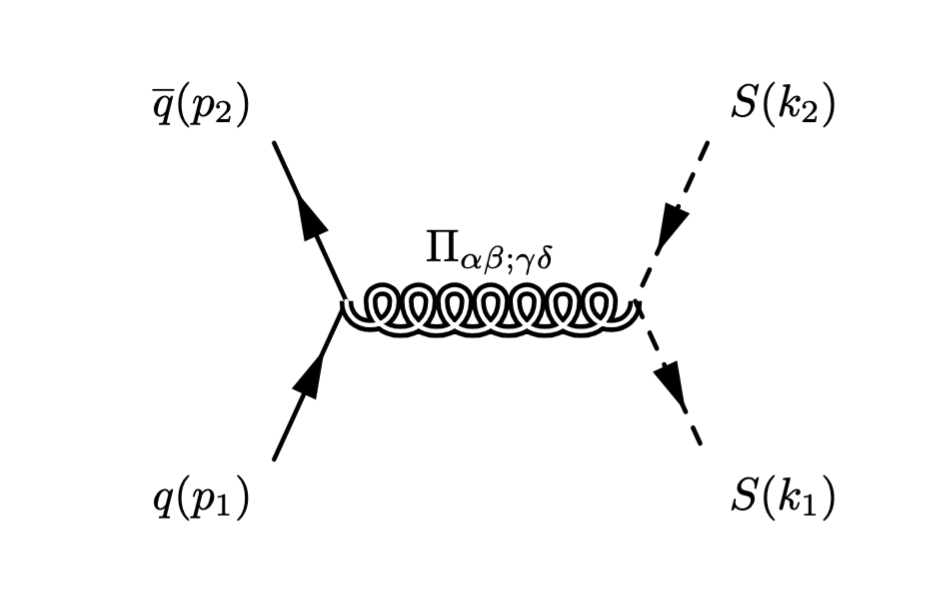}

 \vspace{20pt}
    \caption{Feynman diagram for dark-sector particle production from $\bar q q\to SS$.}
\end{figure}    


We reassess here in detail the estimates of Ref.~\cite{hooman} for constraints from dark-sector particle production in the context of very high-energy cosmic rays. We consider a proton-proton collision at energies comparable to the GZK threshold \cite{Letessier-Selvon:2011sak} leading, at tree level, to gravitational pair production of a pair of dark sector degrees of freedom.  In the center of momentum frame, this corresponds to an $E_{CM} \approx 250 \text{  TeV}$. The relevant conventions are outlined in \cite{FeynGrav}. We  define the  center of mass (c.o.m.) four-momenta for such a scenario as 
\begin{align*}
   & p_1 = (x_1\sqrt{s},0,0,x_1\sqrt{s} ), \hspace{20pt} p_2 = (x_2\sqrt{s},0,0,-x_2\sqrt{s} ), \\[5pt]
&k_1 = (x_1\sqrt{s},0,x_1\sqrt{s} \sin{\theta_1},x_1\sqrt{s} \cos{\theta _1}),   \\[5pt]
&k_2 = (x_2\sqrt{s},0,-x_2\sqrt{s} \sin{\theta_2},-x_2\sqrt{s}\cos{\theta_2} ),
\end{align*}
where the factors $x_1\text{ and }x_2$ denote the fractional energy taken from the proton by a colliding quark and anti-quark. Unless $x_1\text{ and }x_2$ are equivalent, the produced dark particles will scatter at different angles. We can then simplify the phase space integral  for the quark cross section as
\begin{equation}
\begin{split}
&\sigma_{qq} =\\[5pt]
&\frac{1}{x_1 x_2s |v_1-v_2|} \int  \frac{d^3k_1 d^3k_2}{E_1E_2(2 \pi)^2} |\mathcal{M}_{\rm tree}|^2 \delta^4(p_1+p_2-k_1-k_2),
\end{split}
\end{equation}  
with the delta function defined as:
\begin{equation}
\begin{aligned}
&\delta^4(p_1+p_2-k_1-k_2) = \frac{1}{s}\delta_i(0)\delta_j(x_2\sin{\theta_2} -x_1\sin{\theta_1})\\[5pt]
& \times \delta_k(x_1(1 -\cos{\theta_1} )- x_2(1-\cos{\theta_2})) \delta_E(E_{CM}-E_1-E_2).
\end{aligned}
\end{equation}
We  then make the substitution 
\begin{equation*}
\begin{split}
   d^3k_2 = &dk^i_2dk^j_2dk^k_2 \\[5pt]
   = &dk^i_2d(-x_2\sqrt{s}\sin{\theta_2})d(-x_2\sqrt{s}\cos{\theta_2})\\[5pt]
    =&  x_2^2s \cdot dk^i_2d(\sin{\theta_2})d(\cos{\theta_2}).
\end{split}
\end{equation*}
Integrating over delta functions, and making the following substitutions  within the matrix element:
\begin{equation*}
\sin{\theta_2} \rightarrow  \frac{x_1}{x_2}\sin{\theta_1} ,
\end{equation*}
\begin{equation*}
    \cos{\theta_2} \rightarrow  1- \frac{x_1}{x_2}(1-\cos{\theta_1}), 
\end{equation*}
allows us to express $\theta_2$ in terms of fractional energy $x_1,x_2$ and $\theta_1$. Computing using the methods outlined in \cite{FeynGrav} and carrying out substitutions, we find the matrix element:
\begin{equation}
\begin{split}
|\mathcal{M}_{tree}|^2 &=  \frac{\kappa^4 s^2 x_1^2}{32x_2}  [\cos{\theta_1} (x_1 + x_2) - x_1 +  x_2)^2 \\[5pt]
& \times (3 \cos{\theta_1} (x_2 - x_1) + 2 x_1 \cos{2 \theta_1} + x_1 - x_2],
\end{split}
\end{equation}
with  $\kappa^2 = \frac{32 \pi}{M_P^2}$.  

The parton-level cross section equation then becomes 
\begin{equation}
\begin{split}
        \sigma_{qq} =&\frac{1}{x_1 x_2s |v_1-v_2|} \int  \frac{x_2^2 E_1^2 dE_1 d(\cos{\theta_1})d\phi}{E_1E_2(2 \pi)^2} \\[5pt]
        & \hspace{20pt} \times |\mathcal{M}_{Tree}|^2 \delta(E_{CM} -E_1-E_2).
\end{split}
\end{equation}  
where we have integrated over the 3-momentum components of the delta functions with respect to  $dV_2$. We can then integrate $dE_1$ over the remaining delta function to obtain the final integral as a function of $\cos{\theta_1}$. Making the substitution $d(\cos{\theta_1}) \rightarrow dx, \cos{\theta_1} \rightarrow x$ within the matrix element, we get a cross section of 
\begin{equation}
\begin{split}
\sigma_{qq} = \frac{s\kappa^4 x_1}{64\pi }  &\int_{-1}^{1}  \frac{ [E_{CM} -E_2] dx}{ E_2|x_1+x_2|}\bigg{[}((x-1) x_1+(1+x)x_2)^2  \\[5pt]
&\times ((x-1) (4 x+1) x_1+(3 x-1) x_2)\bigg{]},
\end{split}
\end{equation}
where we have simplified $|v_1-v_2| = |x_1+x_2|$ for partons within the c.o.m. frame. Additionally, with  $E_{CM} = \sqrt{x_1x_2s} \text{  and  } E_2 = x_2\sqrt{s}$ , this integral simplifies to 
\vspace{10pt}
\begin{equation}
\sigma_{qq} = \frac{s \kappa^4x_1 (21 x_1^3-23 x_1^2 x_2+x_1 x_2^2+5 x_2^3) (2 \sqrt{x_1 x_2}-x_2)}{240\pi  x_2 |x_1+x_2|},
\end{equation}

\begin{figure*}[t!]
\begin{center}
\includegraphics[width = 0.9\textwidth
]{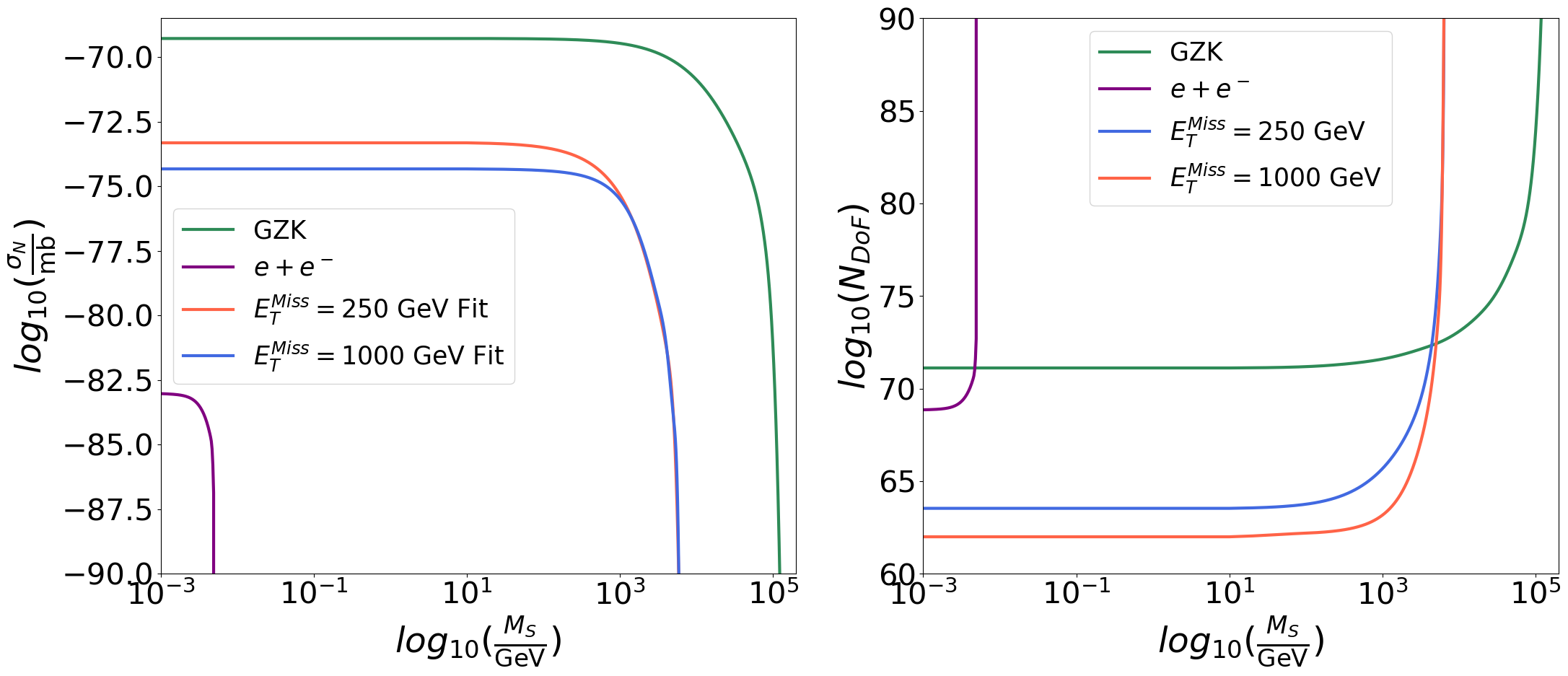}
\caption{ The left panel describes the cross section for select scattering processes for given produced dark sector masses. The lines described by $E_T^{\rm Miss  }$ refer to the LHC constraints outlined in section \ref{sub:LHC}, where $E_T^{\rm Miss  }$ is the missing transverse energy observed by the emitted gluon. The right panel is the computed degrees of freedom via the observed $p \bar{p}$ collision cross section and calculated dark process cross section, i.e. $N \lesssim \sigma_{pp,{\rm tot}}/\sigma_{p p\to SS}$.}.
\label{fig:scattering}
\end{center}
\end{figure*}

    The parton distribution functions (PDFs) we use are those experimentally obtained by the LHAPDF project and  reproduced via the \textit{ManeParse} python package \cite{Mane}. Integrating over the PDFs at $E_{CM} = 250 \text{  
 TeV}$, the final result for our proton- proton cross section is
\begin{equation}
    \sigma_{p p\to SS} =  9.582 \times 10^{-70} \text{ mb}.
\end{equation}
This result informs constraints on the possible number of dark degrees of freedom in the following way: we demand that the {\it total} cross section for gravitationally-produced degrees of freedom be at most equal to the inclusive proton-proton cross section at the same center of mass energy $E_{CM} = 250 \text{  
 TeV}$, $\sigma_{pp,{\rm tot}}(250)\ {\rm TeV}\simeq 100\ {\rm mb}$ \cite{ParticleDataGroup:2024cfk}. As a result, we obtain a maximum number of  degrees of freedom constraint
\begin{equation}
    N \lesssim \sigma_{pp,{\rm tot}}/\sigma_{p p\to SS} \simeq 1.0 \times10^{71}.
\end{equation}
Note that in the estimate above, we neglected the kinematic effect due to the phase-space suppression for dark sector particle masses $M_S$ comparable to half the center of mass energy. When including those effects, the cross section is suppressed, as shown with the green curve in the left panel of fig.~\ref{fig:scattering}. The resulting, dark-sector-mass-dependent constraints are shown with a dark green line in the right panel of the same figure.

We note that in comparison to the order-of-magnitude estimate of $N = 4 \times 10^{68}$ described in Ref \cite{hooman}, our constraints are significantly weaker. As a cross check, using the Randall-Sundrum model implemented in \textit{MadGraph5 aMC@NLO (v2.9.20)} \cite{MadGraph} for comparison, we found  the cross section for the tree level process to be $7.744 \times 10^{-70} \text{ mb}$. This corresponds to $N = 1.291 \times 10^{71} $ degrees of freedom, very close to our value of $ N =  1.043 \times10^{71}$. The relevant Feynman rules for this process are listed in Ref.~\cite{FeynGrav}.

\begin{figure}
    \centering
\includegraphics[width = 0.4\textwidth]{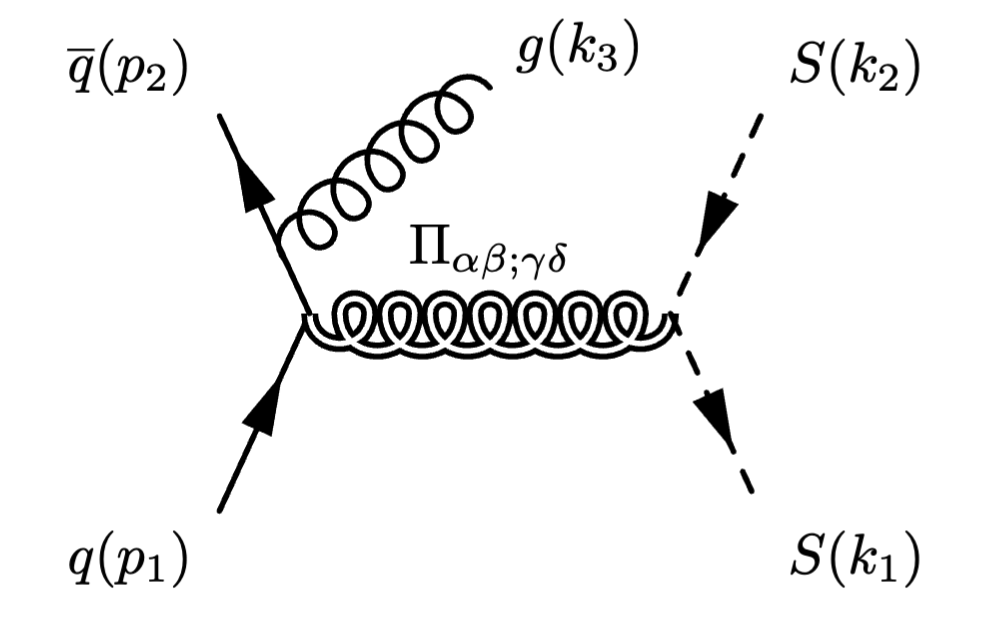}
\vspace{20pt}
    \caption{Feynman diagram relevant for LHC constraints: $\bar q q\to gSS$.}
\end{figure}

\subsection{LHC constraints: Initial State Jet Production with missing transverse energy}\label{sub:LHC}
At colliders, constraints stem from gravitational particle production including an initial-state radiated jet. 
We thus consider an additional initial state gluon radiated for the process described in the section above. Constraints for the number of degrees of freedom will be again obtained via a \textit{MadGraph} analysis of initial state quark and gluon jet  from $p \bar{p}$ collisions at $E_{CM} = 13 \text{ TeV}$. The individual  cross sections will be considerably small due to the gravitational and radiative nature of the process. Still, again, if there are many light species  $N$, the observed cross section at collider experiments will scale as  $N \sigma_{p \bar{p}\to SS+j}$.  The produced dark sector states will be invisible to detectors, which will thus only observe a  monojet with  missing transverse energy/momentum $E_T^{miss}.$ 

We show the results of the cross section in the left panel of fig.~\ref{fig:scattering}. The red line shows the cross section resulting from imposing a cut on $E_T^{miss}>250$ GeV, while the blue line enforces  $E_T^{miss}>1,000$ GeV. The values were purposely chosen to match known experimental results, as discussed in Ref.~\cite{JHEP}. Here, following \cite{hooman}, we consider the total uncertainty on the cross sections for SM processes such as $pp\to \bar \nu \nu$ and on the experimental results, and assume a resulting maximal cross section to non-SM process equivalent to 2\% of the SM's cross section, or $\sim 200$ fb, for $E_T^{miss}>250$ GeV, and of 10\%, or 1 fb, for $E_T^{miss}>1,000$ GeV. The resulting constraints, computed as above as a function of the dark sector common mass scale $M_S$, are shown in the right panel of fig.~\ref{fig:scattering}, with the same color conventions as in the left panel. 

Comparing our detailed estimates to the order-of-magnitude estimate of \cite{hooman}, we find here comparable, albeit, again, slightly weaker results, with $N\lesssim {\rm few}\times 10^{62}$ for light degrees of freedom mass scale $M_S$, versus Ref.~\cite{hooman}'s $N\lesssim 10^{62}$ and Ref.~\cite{Alexeyev:2017scq}'s similar constraints (computed, however, with a different approach).

\subsection{Supernova energy losses}\label{sec:sne}

Supernova energy losses also constrain the existence of a large number of light degrees of freedom (here, the mass needs to be below the typical energy scale of particles radiated in supernova cooling, in the few MeV range). Requiring that the gravitationally produced particles, with a total cross section $\sigma_N\sim NE^2/M_{\rm Pl}^4$, but subdominant compared to neutrino production, approximately characterized by a cross section $\sigma_\nu\sim E^2 G_F^2$, with $G_F$ Fermi's constant, one obtains 
\begin{equation}
N\lesssim G_F^2M_{\rm Pl}^4\sim10^{66}.
\end{equation}

 Using the production cross section obtained above, but for the actual electron-pair-initiated process $e+e^-\to SS$, computed at a center of mass $s\simeq T_{\rm SN}^2\simeq (10\ {\rm MeV})^2$, we find the cross section shown in fig.\ref{fig:scattering}. We obtain constraints on $N$ by requiring the calculated cross section to be smaller than the total production cross section for neutrinos, which can be estimated as \cite{Janka:2017vlw}
 \begin{equation}
\sigma_{\bar\nu\nu}\simeq 1.8\times 10^{-44}\ {\rm cm}^2\left(\frac{T_{\rm SN}}{m_e}\right)^2,     
 \end{equation}
 with $T_{\rm SN}\simeq 10$ MeV \cite{Janka:2017vlw}. Again, computing this value for a massless dark sector we obtain a cross section $ \sigma_{\bar\nu\nu}\simeq  1 \times 10^{-83}\ \text{mb}$. This corresponds to, roughly, $N \simeq 6 \times 10^{68}$ degrees of freedom. The mass dependence of such constraints are shown in the right panel of fig.~\ref{fig:scattering}.

\section{Early universe Gravitational Dark Matter production constraints}\label{sec:cosmo}

A large dark sector can potentially over-close the universe via gravitational particle production in the early universe, assuming the dark degrees of freedom are stable, or that they decay into the lightest particle, which in turn is assumed to be stable. We discuss in this section this possibility.

\begin{figure*}[t!]
\begin{center}
\includegraphics[width = \textwidth, height = .35\textwidth]{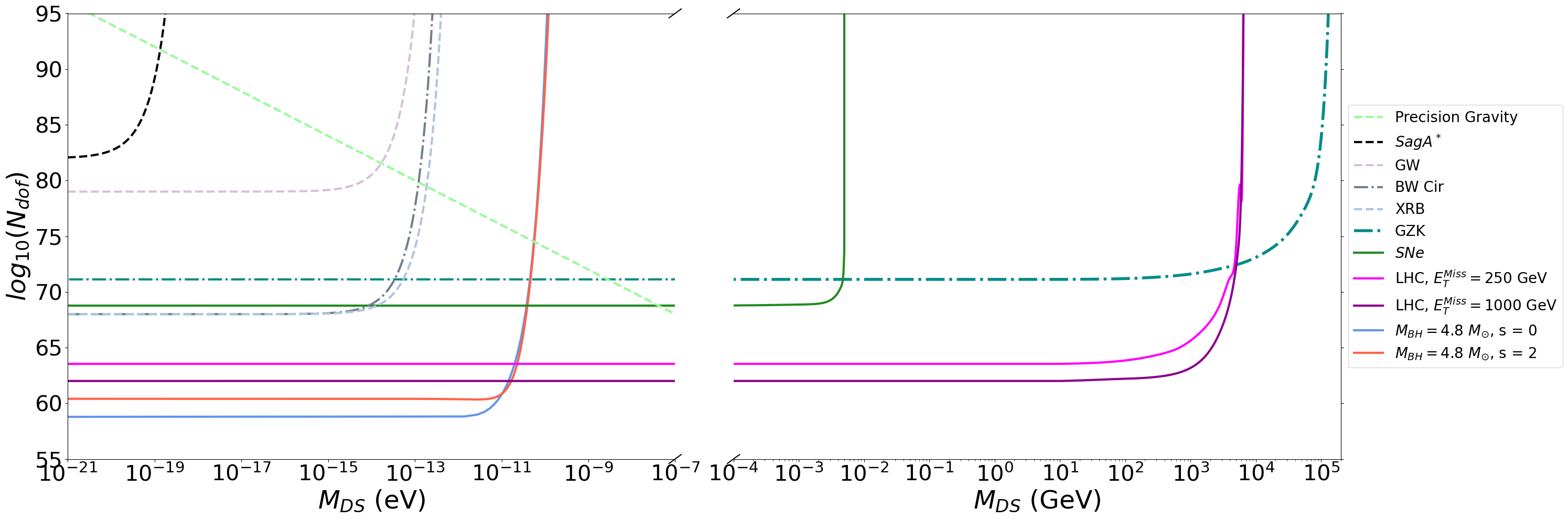}
\caption{Summary plot of all constraints discussed in this study, in the light dark sector region of parameter space (left) and in the heavy sector (right). Parameter space points above each curve are excluded by the corresponding constraint. See the main text for details. }
\label{fig:money}
\end{center}
\end{figure*}

Using  results from Ref.~\cite{Mambrini},  the fraction of critical density of dark matter per degree of freedom can be expressed as 
\begin{equation}
    \begin{split}
        \frac{\Omega_S h^2}{0.1} \approx \left(\frac{T_{RH}}{10^{10} \text{GeV}}\right)^3\left(\frac{T_{\rm max}/T_{RH}}{100}\right)^4\\[0.2cm]
        \left(\frac{m_S}{1.75 \times 10^{10}\text{GeV}}\right) \left(1+ \frac{m_S^2}{2m_{\phi}^2}\right)\sqrt{1- \frac{m_S^2}{m_{\phi}^2}},
    \end{split}
\end{equation}
where $\phi$ and $\textit{S}$ refer to the inflaton and dark sector species, respectively, and $T_{\rm max}>T_{RH}$ is the initial temperature of the radiation bath relaxing the instantaneous reheating approximation \cite{Mambrini}. Assuming $m_S \ll m_{\phi}$, this reduces to 
\begin{equation}
\frac{\Omega_S h^2}{0.1} \approx \left(\frac{T_{RH}}{10^{10} \text{GeV}}\right)^3\left(\frac{T_{\rm max}/T_{RH}}{100}\right)^4\left(\frac{m_S}{1.75 \times 10^{10} \text{GeV}}\right).
\end{equation}

Similarly, Ref.~\cite{Ema} showed a similar result in their paper; assuming minimal coupling, they find
\begin{equation}
   \frac{\Omega_S h^2}{0.1} \approx 2
     \mathcal{C}\  \left(\frac{m_S}{m_\phi}\right)^4\left(\frac{m_S}{10^9 \ \text{GeV}}\right)\left(\frac{H_{inf}}{10^9\ \text{GeV}}\right)\left(\frac{T_{RH}}{10^{10}\ \text{GeV}}\right). 
\end{equation} 
which describes gravitational production, and where the coefficient  $\mathcal{C}$ has been computed numerically to be of order $10^{-3} - 10^{-2}$ \cite{Ema}. 
Additionally, in the early universe the thermal bath at temperature $T$ unavoidably leads to gravitational particle production via ${\rm SM}+{\rm SM}\to SS$, with a cross section that can be estimated as 
\begin{equation}
    \sigma_{{\rm SM}+{\rm SM}\to SS}\sim {\cal A}\frac{T^2}{M_{\rm Pl}^4}.
\end{equation}
and ${\cal A}\sim {\rm few}\times (1/(12\pi))$ depending on the particles in the thermal bath and their spin. The resulting ``thermal'' production yields
\begin{eqnarray}\nonumber
    \frac{\Omega^{\rm th}_S h^2}{0.1} &\approx&   10^{-15} \left(\frac{\cal A}{200}\right)\left(\frac{m_S}{10^9 \ \text{GeV}}\right)\\
    &&\times \left(\frac{10^9\ \text{GeV}}{H_{inf}}\right)\left(\frac{T_{RH}}{10^{10}\ \text{GeV}}\right)^4. 
\end{eqnarray}
Finally, for light dark sectors, coherent oscillation could also lead to significant gravitational production \cite{Ema}.

Depending on inflation models, the constraints described above would imply very stringent bounds on the number of {\it stable} dark degrees of freedom. However, as mentioned above, if the dark sector is not stable, such constraints do not apply.

If, however, the dark sector is stable, we find that the constraints on $N$ from overproduction in the early universe via gravitational and thermal channels read, for the nominal values for $T_{\rm RH}\ T_{\rm max},\ {\cal A},\ {\rm and}\ H_{\rm inf}$ used in the equations above:

\begin{equation}
    N_{\rm GP}\lesssim \frac{1.75 \times 10^{10}\  \text{GeV}}{m_S},
\end{equation}
\begin{equation}
    N_{\rm th}\lesssim \frac{10^{-15}\  \text{GeV}}{m_S},
\end{equation}
We will not show the constraints in our final plot as they apply to a very different range of masses and degrees of freedom than the other constraints described above, and rely upon different assumptions.

\section{Discussion and Conclusions}\label{sec:conclusions}

In this paper, we reassessed in detail model-independent constraints on a large dark sector consisting of purely gravitationally interacting particles. A major caveat to our analysis pertains to constraints from strong gravity near the electroweak scale, as discussed in the introduction. If those are avoided -- and we presented a few plausible reasons why this is not unlikely -- constraints from renormalization group running of the gravitational coupling are weaker than two generic classes of constraints.

The first class pertains to the evaporation of stellar-mass BHs: while with a number of degrees of freedom similar to the that of the SM of particle physics, the Hawking evaporation rate of stellar-mass BHs is extremely small, in the presence of a very large number of additional degrees of freedom the lifetime of a stellar-mass BH could be well shorter than the age of the universe, and potentially of the age of the BH itself as well. This statement depends both on the spin of the BH and on the spin and mass of the dark sector degrees of freedom. We studied in detail how LIGO/VIRGO data on gravitational wave from BH mergers, informing on the mass and spin of the BHs, provide the strongest constraints on the evaporation rate and studied the aforementioned effects from the dark sector particle properties on the constraints on the size of such dark sector.

We also discussed how BH constraints could be avoided assuming the mass of the BHs at production was so large that the BHs could accommodate a very large evaporation rate and have the observed mass at the ``right time''. In this case, however, we were able to set weaker constraints from a number of considerations pertaining to direct effects of the {\it current} evaporation rate (on merging BHs, on the luminosity variation of X-ray binary systems, and on indirect BH mass measurements over time).

The second class of constraints pertains to the linearly enhanced cross section, with the number of dark degrees of freedom, for gravitational particle production (i.e. graviton-mediated production). Three constraints are relevant here: production at very high-energy cosmic ray events, where these processes would compete for instance with the observed GZK cutoff; production, in conjunction with initial state radiation, at the LHC; and production, with ensuing cooling effects, at supernova explosions. All these constraints are weaker that those from evaporation, but extend to significantly more massive dark sectors.

Finally, we discussed gravitational and thermal dark sector particle production. Depending on model-dependent assumptions, these constraints can be extremely tight, but they disappear if the dark sector particles are metastable.

In our final fig.~\ref{fig:money} we show all of the constraints discussed in this study, on the plane defined by, the number of dark degrees of freedom $N_{\rm dof}$ versus the mass of the dark sector particles, all assumed to be degenerate at a  mass $M_{\rm DS}$. The left panel focuses on the low-mass end, where BH evaporation provides the tightest constraints. We show results for both spin 0 and spin 2 dark sectors with the blue and red curves, respectively. The right panel of the figure refers to the heavy sector and illustrates the mass dependence of scattering processes from the LHC, supernovae and cosmic rays. We also show the constraints from precision gravity.

Comparisons to measured proton-proton cross sections within LHC collisions offer the tightest constraints across a vast mass range, and are stronger than the weaker version of constraints from BH evaporation. Notice that, if extended to the large-mass range, precision gravity constraints would rule out the entirety of the right panel - but, as pointed out before, such constraints come with caveats. 

In the most conservative-possible scenario of a large dark sector populated by spin 2 particles, the maximal number of degrees of freedom is $N_{\rm max,massless}< 2.5 \times 10^{60}$.

\section*{Acknowledgements}
This work is partly supported by the U.S.\ Department of Energy grant number de-sc0010107 (SP).

\bibliographystyle{unsrt} 
\bibliography{bibliography}

\begin{thebibliography}{10}

\bibitem{Chacko:2005pe}
Z.~Chacko, Hock-Seng Goh, and Roni Harnik.
\newblock {The Twin Higgs: Natural electroweak breaking from mirror symmetry}.
\newblock {\em Phys. Rev. Lett.}, 96:231802, 2006.

\bibitem{Martin:1997ns}
Stephen~P. Martin.
\newblock {A Supersymmetry primer}.
\newblock {\em Adv. Ser. Direct. High Energy Phys.}, 18:1--98, 1998.

\bibitem{Foot:1991bp}
Robert Foot, H.~Lew, and R.~R. Volkas.
\newblock {A Model with fundamental improper space-time symmetries}.
\newblock {\em Phys. Lett. B}, 272:67--70, 1991.

\bibitem{Arkani-Hamed:1998jmv}
Nima Arkani-Hamed, Savas Dimopoulos, and G.~R. Dvali.
\newblock {The Hierarchy problem and new dimensions at a millimeter}.
\newblock {\em Phys. Lett. B}, 429:263--272, 1998.

\bibitem{Hooper:2007qk}
Dan Hooper and Stefano Profumo.
\newblock {Dark Matter and Collider Phenomenology of Universal Extra Dimensions}.
\newblock {\em Phys. Rept.}, 453:29--115, 2007.

\bibitem{Giudice:2016yja}
Gian~F. Giudice and Matthew McCullough.
\newblock {A Clockwork Theory}.
\newblock {\em JHEP}, 02:036, 2017.

\bibitem{Dienes:2021woi}
Keith~R. Dienes, Lucien Heurtier, Fei Huang, Doojin Kim, Tim M.~P. Tait, and Brooks Thomas.
\newblock {Stasis in an expanding universe: A recipe for stable mixed-component cosmological eras}.
\newblock {\em Phys. Rev. D}, 105(2):023530, 2022.

\bibitem{ref17}
Xavier Calmet, Stephen D.~H. Hsu, and David Reeb.
\newblock {Quantum gravity at a TeV and the renormalization of Newton's constant}.
\newblock {\em Phys. Rev. D}, 77:125015, 2008.

\bibitem{ref47}
Pierre Fayet.
\newblock {MICROSCOPE limits for new long-range forces and implications for unified theories}.
\newblock {\em Phys. Rev. D}, 97(5):055039, 2018.

\bibitem{ref48}
Stephan Schlamminger, K.~Y. Choi, T.~A. Wagner, J.~H. Gundlach, and E.~G. Adelberger.
\newblock {Test of the equivalence principle using a rotating torsion balance}.
\newblock {\em Phys. Rev. Lett.}, 100:041101, 2008.

\bibitem{Calmet}
Xavier Calmet, Stephen D.~H. Hsu, and David Reeb.
\newblock Quantum gravity at a tev and the renormalization of newton’s constant.
\newblock {\em Physical Review D}, 77(12), June 2008.

\bibitem{ref19}
Adri\'an del Rio, Ruth Durrer, and Subodh~P. Patil.
\newblock {Tensor Bounds on the Hidden Universe}.
\newblock {\em JHEP}, 12:094, 2018.

\bibitem{Pal:2019tqq}
K.~Pal, L.~V. Sales, and J.~Wudka.
\newblock {Ultralight Thomas-Fermi dark matter}.
\newblock {\em Phys. Rev. D}, 100(8):083007, 2019.

\bibitem{Hawking}
G.~W. Gibbons and S.~W. Hawking.
\newblock Cosmological event horizons, thermodynamics, and particle creation.
\newblock {\em Phys. Rev. D}, 15:2738--2751, May 1977.

\bibitem{BH}
Alexandre Arbey and Jérémy Auffinger.
\newblock Blackhawk: a public code for calculating the hawking evaporation spectra of any black hole distribution.
\newblock {\em The European Physical Journal C}, 79(8), August 2019.

\bibitem{Page:1976ki}
Don~N. Page.
\newblock {Particle Emission Rates from a Black Hole. 2. Massless Particles from a Rotating Hole}.
\newblock {\em Phys. Rev. D}, 14:3260--3273, 1976.

\bibitem{hooman}
Hooman Davoudiasl, Peter~B. Denton, and David~A. McGady.
\newblock Ultralight fermionic dark matter.
\newblock {\em Physical Review D}, 103(5), March 2021.

\bibitem{Ligo}
Alexander~H. Nitz, Sumit Kumar, Yi-Fan Wang, Shilpa Kastha, Shichao Wu, Marlin Sch\"afer, Rahul Dhurkunde, and Collin~D. Capano.
\newblock {4-OGC: Catalog of Gravitational Waves from Compact Binary Mergers}.
\newblock {\em Astrophys. J.}, 946(2):59, 2023.

\bibitem{BHGW150914}
B.~P.~\textit{et al} Abbott.
\newblock The basic physics of the binary black hole merger gw150914.
\newblock {\em Annalen der Physik}, 529(1–2), October 2016.

\bibitem{Formr}
Razieh Emami and Abraham Loeb.
\newblock Formation redshift of the massive black holes detected by ligo, 2019.

\bibitem{Kalogera:1996ci}
Vassiliki Kalogera and Gordon Baym.
\newblock {The maximum mass of a neutron star}.
\newblock {\em Astrophys. J. Lett.}, 470:L61--L64, 1996.

\bibitem{Fernandez_2019}
Nicolas Fernandez and Stefano Profumo.
\newblock Unraveling the origin of black holes from effective spin measurements with ligo-virgo.
\newblock {\em Journal of Cosmology and Astroparticle Physics}, 2019(08):022–022, August 2019.

\bibitem{discharge}
Jos\'e~Antonio de~Freitas~Pacheco.
\newblock {Evaporation of Primordial Charged Black Holes: Timescale and Evolution of Thermodynamic Parameters}.
\newblock {\em Symmetry}, 16(7):895, 2024.

\bibitem{Kerr}
Roy~P. Kerr.
\newblock Gravitational field of a spinning mass as an example of algebraically special metrics.
\newblock {\em Phys. Rev. Lett.}, 11:237--238, Sep 1963.

\bibitem{Arbey:2021mbl}
Alexandre Arbey and J\'er\'emy Auffinger.
\newblock {Physics Beyond the Standard Model with BlackHawk v2.0}.
\newblock {\em Eur. Phys. J. C}, 81:910, 2021.

\bibitem{carroll}
Sean~M. Carroll.
\newblock {\em {Spacetime and Geometry}: {An Introduction to General Relativity}}.
\newblock Cambridge University Press, 7 2019.

\bibitem{2009ApJS..181..238C}
J.~{Casares}, J.~A. {Orosz}, C.~{Zurita}, T.~{Shahbaz}, J.~M. {Corral-Santana}, J.~E. {McClintock}, M.~R. {Garcia}, I.~G. {Mart{\'\i}nez-Pais}, P.~A. {Charles}, R.~P. {Fender}, and R.~A. {Remillard}.
\newblock {Refined Orbital Solution and Quiescent Variability in the Black Hole Transient GS 1354-64 (= BW Cir)}.
\newblock {\em Ap. J. S.}, 181(1):238--243, March 2009.

\bibitem{Filippenko:1999zv}
A.~V. Filippenko, D.~C. Leonard, T.~Matheson, W.~Li, E.~C. Moran, and A.~G. Riess.
\newblock {A black hole in the x-ray nova velorum 1993}.
\newblock {\em Publ. Astron. Soc. Pac.}, 111:969--979, 1999.

\bibitem{Plotkin:2021rzl}
R.~M. Plotkin, A.~Bahramian, J.~C.~A. Miller-Jones, M.~T. Reynolds, P.~Atri, T.~J. Maccarone, A.~W. Shaw, and P.~Gandhi.
\newblock {Towards a larger sample of radio jets from quiescent black hole X-ray binaries}.
\newblock {\em Mon. Not. Roy. Astron. Soc.}, 503(3):3784--3795, 2021.

\bibitem{GRAVITY:2021xju}
R.~Abuter et~al.
\newblock {Mass distribution in the Galactic Center based on interferometric astrometry of multiple stellar orbits}.
\newblock {\em Astron. Astrophys.}, 657:L12, 2022.

\bibitem{Letessier-Selvon:2011sak}
Antoine Letessier-Selvon and Todor Stanev.
\newblock {Ultrahigh Energy Cosmic Rays}.
\newblock {\em Rev. Mod. Phys.}, 83:907--942, 2011.

\bibitem{FeynGrav}
B.~Latosh.
\newblock Feyngrav 2.0.
\newblock {\em Computer Physics Communications}, 292:108871, November 2023.

\bibitem{Mane}
D.B. Clark, E.~Godat, and F.I. Olness.
\newblock Maneparse: A mathematica reader for parton distribution functions.
\newblock {\em Computer Physics Communications}, 216:126–137, July 2017.

\bibitem{ParticleDataGroup:2024cfk}
S.~Navas et~al.
\newblock {Review of particle physics}.
\newblock {\em Phys. Rev. D}, 110(3):030001, 2024.

\bibitem{MadGraph}
J.~Alwall, R.~Frederix, S.~Frixione, V.~Hirschi, F.~Maltoni, O.~Mattelaer, H.-S. Shao, T.~Stelzer, P.~Torrielli, and M.~Zaro.
\newblock The automated computation of tree-level and next-to-leading order differential cross sections, and their matching to parton shower simulations.
\newblock {\em Journal of High Energy Physics}, 2014(7), July 2014.

\bibitem{JHEP}
M.~\textit{et al} Aaboud.
\newblock Search for dark matter and other new phenomena in events with an energetic jet and large missing transverse momentum using the atlas detector.
\newblock {\em Journal of High Energy Physics}, 2018(1), January 2018.

\bibitem{Alexeyev:2017scq}
S.~O. Alexeyev, X.~Calmet, and B.~N. Latosh.
\newblock {Gravity induced non-local effects in the standard model}.
\newblock {\em Phys. Lett. B}, 776:111--114, 2018.

\bibitem{Janka:2017vlw}
H.~Th. Janka.
\newblock {Neutrino Emission from Supernovae}.
\newblock 2 2017.

\bibitem{Mambrini}
Yann Mambrini and Keith~A. Olive.
\newblock Gravitational production of dark matter during reheating.
\newblock {\em Physical Review D}, 103(11), June 2021.

\bibitem{Ema}
Yohei Ema, Kazunori Nakayama, and Yong Tang.
\newblock Production of purely gravitational dark matter.
\newblock {\em Journal of High Energy Physics}, 2018(9), September 2018.

\end{thebibliography}

\end{document}